\documentclass[twoside,twocolumn,9pt]{article}
\usepackage{extsizes}
\usepackage[super,sort&compress,comma]{natbib} 
\usepackage[version=3]{mhchem}
\usepackage[left=1.5cm, right=1.5cm, top=1.785cm, bottom=2.0cm]{geometry}
\usepackage{balance}
\usepackage{times,mathptmx}
\usepackage{sectsty}
\usepackage{graphicx} 
\usepackage{lastpage}
\usepackage[format=plain,justification=justified,singlelinecheck=false,font={stretch=1.125,small,sf},labelfont=bf,labelsep=space]{caption}
\usepackage{float}
\usepackage{fancyhdr}
\usepackage{fnpos}
\usepackage[english]{babel}
\addto{\captionsenglish}{%
  
}
\usepackage{array}
\usepackage{droidsans}
\usepackage{charter}
\usepackage[T1]{fontenc}
\usepackage[usenames,dvipsnames]{xcolor}
\usepackage{setspace}
\usepackage[compact]{titlesec}
\usepackage{hyperref}

\usepackage{epstopdf}

\newcommand{\be}{\begin{equation}} 
\newcommand{\ee}{\end{equation}}
\newcommand{\bea}{\begin{eqnarray}}   
\newcommand{\eea}{\end{eqnarray}}

\newcommand{\xx}{\mathbf{x}}
\newcommand{\vv}{\mathbf{v}}
\newcommand{\uu}{\mathbf{u}}

\definecolor{cream}{RGB}{222,217,201}

\begin{document}

\pagestyle{fancy}
\thispagestyle{plain}
\fancypagestyle{plain}{

\renewcommand{\headrulewidth}{0pt}
}

\makeFNbottom
\makeatletter
\renewcommand\LARGE{\@setfontsize\LARGE{15pt}{17}}
\renewcommand\Large{\@setfontsize\Large{12pt}{14}}
\renewcommand\large{\@setfontsize\large{10pt}{12}}
\renewcommand\footnotesize{\@setfontsize\footnotesize{7pt}{10}}
\makeatother

\renewcommand{\thefootnote}{\fnsymbol{footnote}}
\renewcommand\footnoterule{\vspace*{1pt}%
\color{cream}\hrule width 3.5in height 0.4pt \color{black}\vspace*{5pt}} 
\setcounter{secnumdepth}{5}

\makeatletter 
\renewcommand\@biblabel[1]{#1}            
\renewcommand\@makefntext[1]%
{\noindent\makebox[0pt][r]{\@thefnmark\,}#1}
\makeatother 
\renewcommand{\figurename}{\small{Fig.}~}
\sectionfont{\sffamily\Large}
\subsectionfont{\normalsize}
\subsubsectionfont{\bf}
\setstretch{1.125} 
\setlength{\skip\footins}{0.8cm}
\setlength{\footnotesep}{0.25cm}
\setlength{\jot}{10pt}
\titlespacing*{\section}{0pt}{4pt}{4pt}
\titlespacing*{\subsection}{0pt}{15pt}{1pt}

\fancyfoot{}
\fancyfoot[LO,RE]{\vspace{-7.1pt}\includegraphics[height=9pt]{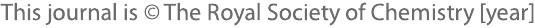}}
\fancyfoot[CO]{\vspace{-7.1pt}\hspace{13.2cm}\includegraphics{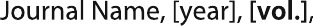}}
\fancyfoot[CE]{\vspace{-7.2pt}\hspace{-14.2cm}\includegraphics{head_foot/RF}}
\fancyfoot[RO]{\footnotesize{\sffamily{1--\pageref{LastPage} ~\textbar  \hspace{2pt}\thepage}}}
\fancyfoot[LE]{\footnotesize{\sffamily{\thepage~\textbar\hspace{3.45cm} 1--\pageref{LastPage}}}}
\fancyhead{}
\renewcommand{\headrulewidth}{0pt} 
\renewcommand{\footrulewidth}{0pt}
\setlength{\arrayrulewidth}{1pt}
\setlength{\columnsep}{6.5mm}
\setlength\bibsep{1pt}

\makeatletter 
\newlength{\figrulesep} 
\setlength{\figrulesep}{0.5\textfloatsep} 

\newcommand{\topfigrule}{\vspace*{-1pt}%
\noindent{\color{cream}\rule[-\figrulesep]{\columnwidth}{1.5pt}} }

\newcommand{\botfigrule}{\vspace*{-2pt}%
\noindent{\color{cream}\rule[\figrulesep]{\columnwidth}{1.5pt}} }

\newcommand{\dblfigrule}{\vspace*{-1pt}%
\noindent{\color{cream}\rule[-\figrulesep]{\textwidth}{1.5pt}} }

\makeatother

\twocolumn[
  \begin{@twocolumnfalse}
\sffamily
\begin{tabular}{m{1.0cm} p{15.cm} }

& \noindent\LARGE{\textbf{Active chiral particles under confinement: \,\,\, \,\,\, \,\,\,
 \,\,\, \,\,\, \,\,\, \,\,\,surface currents and bulk accumulation phenomena}}\\
\vspace{0.3cm} & \vspace{0.3cm} \\

 & \noindent\large{Lorenzo Caprini$^{\ast}$\textit{$^{a}$} and Umberto Marini Bettolo Marconi \textit{$^{b}$}} \\

& \noindent\normalsize{In this work, we study the stationary behavior of an assembly of independent chiral active particles under confinement by
employing an extension of the active Ornstein-Uhlenbeck model. The chirality modeled
by means of an effective torque term leads to a drastic reduction of the accumulation near the walls with respect to the 
case without handedness and to the appearance of currents parallel to the container walls
accompanied by a large accumulation of particles in the inner region.
In the case of two-dimensional chiral particles confined by harmonic walls, we determine the analytic form of the distribution of positions and velocities in two different situations: a rotationally invariant
confining potential and an infinite channel with parabolic walls. Both these models display currents and chirality induced inner accumulation. 
These phenomena are further investigated by means of a more realistic description of a channel, where the wall and bulk regions are clearly separated. The corresponding current and density profiles are obtained by numerical simulations.
At variance with the harmonic models, the third model shows a progressive emptying of the wall regions 
and the simultaneous enhancement of the bulk population.
We explain such a phenomenology in terms of the combined effect of wall repulsive forces and chiral motion and
provide a semiquantitative description of the current profile in terms of an effective viscosity of the chiral
gas.}

\end{tabular}

 \end{@twocolumnfalse} \vspace{0.6cm}

  ]

\renewcommand*\rmdefault{bch}\normalfont\upshape
\rmfamily
\section*{}
\vspace{-1cm}


\footnotetext{\textit{$^{a}$~Gran Sasso Science Institute (GSSI), Via. F. Crispi 7, 67100 L'Aquila, Italy. E-mail: lorenzo.caprini@gssi.it}}
\footnotetext{\textit{$^{b}$School of Sciences and Technologies, Universit\`a di Camerino, via Madonna delle Carceri, 62032, Camerino, and INFN Perrugia, Italy. 
E-mail: umberto.marinibettolo@unicam.it  }}




\section{Introduction}
The dynamics of self-propelled particles is an exciting new area of research at the frontier between physics, biology and bioengineering
\cite{ramaswamy2010mechanics,bechinger2016active,marchetti2013hydrodynamics,romanczuk2012active}.These particles are ubiquitous in nature or can be artificially created and employed in a near future in tasks such as targeted drug delivery and nanosurgery. Examples of the first category are bacteria, ciliates, Synechococcus a type of blue-green alga, sperm cells, while artificial self-propellers are represented by Janus particles catalytically driven or magnetic microparticles subject to a magnetic field. 
 Each particle, whose size may range between 10 $\mu m$ up to  several microns, consumes energy from the environment in order to move and is fueled either either by metabolic processes or by chemical reactions: E. coli move forward by a rotational motion of their spiral-shaped flagella while synthetic Janus colloids are driven by  a catalytic chemical reaction on their surface. 
  
It is sufficient a small
departure from the right-left symmetry relative to the propulsion axis of self-propelled particles
or an external magnetic field to determine circle swimming in two dimensions or helical swimming in three dimensions, a phenomenon termed as chirality or handedness. Indeed, in nature one observes spiral-like swimming trajectories in E. coli bacteria~\cite{lauga2006swimming} and spermatozoa in bulk suspensions~\cite{friedrich2008stochastic} or circle-like motion near a planar substrate~\cite{diluzio2005escherichia} and  clockwise treadmilling in FtsZ proteins on membranes~\cite{loose2014bacterial}.

Analogous chiral trajectories are produced by artificial microswimmers, for instance, L-shaped particles \cite{kummel2013circular,lowen2016chirality}.
In nature, magnetotactic bacteria are widespread motile prokaryotes, endowed with an organelle
called magnetosome which causes the cell to behave like a miniature compass and swim parallel to the magnetic field lines ~\cite{blakemore1975magnetotactic,lefevre2013ecology}. The direction of the magnetic field is so crucial that this kind of bacteria die when taken to the opposite hemisphere of Earth.

 As discussed by ten Hagen et al. \cite{ten2015can} a system of self-propelled particles can be described at two different
levels: i) a fine-grained level fully taking into account the self-propulsion mechanism, the degrees of freedom of the microswimmers, such as moving internal elements,
and the solvent hydrodynamics  \cite{najafi2004simple}, ii) a coarse-grained level where 
the motion of  microswimmers is modeled by using the Brownian overdamped equations
and the self-propulsion is represented
in terms of effective forces and torques~\cite{van2008dynamics}.
In the following, we have found convenient, following the literature in the field, to adopt
the second level of description which allows a straightforward
application of the tools of statistical mechanics and computer simulation to large collections of active particles. Among these tools there are
two popular theoretical approaches to the study of active particles: 
Active Brownian
particle (ABP) model~\cite{ebeling1999active,cates2013active}, and the Active Ornstein-Uhlenbeck particle (AOUP) model~\cite{szamel2014self}. Both models represent the microswimmer motion through a viscous
solvent by means of Langevin overdamped dynamics and include the distinguishing feature of active particles, namely
the self-propulsion mechanism as an effective force term.
Such a self-propulsion force has the following properties: a) it is typically noisy
and its orientation is random, b) active forces acting on different self-propelled particles are mutually independent and c)  the direction of driving persists over time scales
of the order of the microseconds.
As briefly discussed in the next section,
the ABP and AOUP differ by the way the self-propulsion force is modeled, but in spite of that
many important aspects of their nonequilibrium behavior are similar.
So far, the mechanism responsible for the active chiral behavior has been included only in the ABP  by adding an effective constant torque term in the corresponding Langevin equation~\cite{van2008dynamics}.
   Such a torque induces a systematic
drift in the orientation of active particles, i.e. a rotation with a constant angular speed of the active velocity, leading to helical trajectories in three dimensions and to circular ones in two dimensions.
While the effect of chirality on the bulk behavior of self-propelled particles has been investigated \cite{lowen2016chirality}, 
the properties of self-propelled chiral particles under confinement are not so well known and understood.
Confining surfaces may induce surface accumulation, sliding motion and selection of active particles according to their chirality~\cite{mijalkov2013sorting,volpe2014simulation,li2014manipulating} and therefore 
it is interesting to investigate under which conditions these effects can be enhanced or suppressed.

In the present work, we focus on the properties of confined chiral particles and motivated by the simpler mathematical structure of the AOUP, which facilitates
the theoretical treatments, we extend it to account for chiral effects. 
As in the case of the ABP,  this goal is achieved by adding a torque term to the self-propulsion.

The paper is organized as follows: In Sec. \ref{chiralmodel} we introduce the chiral active Ornstein-Uhlenbeck particle (CAOUP) model, in Sec.~\ref{circular} we study the special case of CAOUP confined in a two-dimensional harmonic trap,
in Sec.~\ref{confinement} we study the behaviour of the model in the case of confinement in a parabolic channel and in a slit.
 Finally, in Sec~\ref{conclusions} we present the conclusions and discuss some future directions.
  
 \section{ Planar chiral active motion}
 \label{chiralmodel}
 
In the following, we shall describe an assembly of $N$ mutually independent active chiral  particles moving in the
two dimensional $(x,y)$ plane. Each particle at position ${\mathbf{x}}_i$
is subject to four kinds of drivings:
 an active or self-propulsion force,  $ {\mathbf f}^a_i  $, a drag force, $-\gamma \dot{\mathbf{x}}_i$  
due to the friction with the solvent, a white noise force $\gamma \sqrt{2D_t}\boldsymbol{\xi}_i$ representing the thermal agitation of the solvent, whose intensity depends on the thermal diffusion coefficient, $D_t$,
and  a force, $-\nabla\phi( {\mathbf x }_i)$ due to a potential, $\phi$ which confines
their motion in a restricted region of the two-dimensional space.
The active force, $ {\mathbf f}^a_i  =\gamma v_0 {\mathbf{e}}_i $ has a constant magnitude proportional to 
a velocity $ v_0$ and
a direction $ {\mathbf{e}}_i\equiv (\cos\theta_i,\sin\theta_i)$ dependent on an
 angle $\theta_i(t)$ undergoing unbiased rotational diffusion with rotational diffusion coefficient  $D_r$.
Inertial effects are neglected because of the low Reynolds number regime as well as
hydrodynamic effects and the solvent reaction. 
Van Teffelen et al.~\cite{van2008dynamics} have extended the standard ABP model to account for
the chiral motion of microswimmers by adding a new ingredient: they imposed a constant angular drift of amplitude $\Omega$ to the dynamics of the angle $\theta_i(t)$ to represent an effective constant torque uniformly applied to the particles.
 The resulting chiral ABP model is described by the following stochastic equations:
  \begin{eqnarray}
\label{eq:ABPdynamics1}
\gamma\dot {\mathbf x }_i &=& -\nabla\phi( {\mathbf x }_i) + \gamma\sqrt{2D_t}\boldsymbol{\xi}_i + {\mathbf{f}}^a_i , \\
\dot{\mathbf{e}}_i &=&(\sqrt{2D_r}w_i+ \Omega) \hat{ \mathbf{z}} \times \mathbf{e}_i  ,
\label{eq:ABPdynamics2}
\end{eqnarray}
 where  $\hat{ \mathbf{z}}$ is the unit vector normal to such a plane, $\boldsymbol{\xi}_i$ and $w_i$ are independent Gaussian noises with $\delta$-correlated components, unit variance and zero mean. 
The torque turns the standard exponential form of the autocorrelation function of the orientation vector
into a damped oscillatory behavior
\begin{equation}
\label{eq:unitvector}
\begin{aligned}
\langle  {\mathbf{e}}_i(t)\cdot   {\mathbf{e}}_j   (0) \rangle= \delta_{ij}\, e^{-  D_r t }\, \cos(\Omega
t) ,
\end{aligned}
\end{equation}
where $t>0$. Notice that when  $|\Omega t |>\pi/2$ the force autocorrelation function 
becomes negative.

A series of
interesting studies of the chiral ABP have been presented by various authors \cite{mijalkov2013sorting,li2014manipulating,ai2015chirality,ao2015diffusion},  
and mainly rely on numerical simulations and/or the analysis of the low noise, quasi-deterministic behavior of
equations \eqref{eq:ABPdynamics1} and \eqref{eq:ABPdynamics2}.

Hereafter, motivated by the success of the active Ornstein-Uhlenbeck model in reproducing
and predicting the main behaviors of non-chiral microswimmers~\cite{marconi2015towards,das2018confined,caprini2018activity}, we consider its chiral extension.
Moreover, the AOUP is considered to be a valid alternative tool to investigate the properties of active particles because of the feature which makes it analytically more treatable than  the ABP, namely
the property that the fluctuations of the self-propulsion force are Gaussian.
We introduce the chiral version of the AOUP model by 
assuming the same governing equation \eqref{eq:ABPdynamics1}  for ${\mathbf{x}}_i$
as in the ABP, but writing
the dynamics of the active force ${\mathbf{f}^a_i}=\gamma{\mathbf{u}_i}$  as:
\begin{equation}
\dot{\mathbf{u}}_i=  - \frac{\mathbf{u}_i}{\tau} +\Omega \hat{ \mathbf{z}} \times \mathbf{u}_i  + \frac{\sqrt{2D_a }}{\tau}\boldsymbol{\eta}_i,
\label{AOUPdyn}
\end{equation}
where $\mathbf{\eta}_i$ is a Gaussian noise and each component of $\uu_i$ evolves according
to an Ornstein-Uhlenbeck  process of characteristic time, $\tau$ and strength $D_a$ and is subject to a tangential drift at a fixed frequency $\Omega$ around an axis orthogonal to the plane of motion.
It is very simple to show that, in the case of freely moving particles ($\phi=0$),  the two-time correlation functions of the chiral versions of ABP and AOUP
are similar.
A brief calculation gives the following two-time the average:
\be
 \langle u_\alpha(t) u_\alpha(0)\rangle = \frac{D_a}{\tau} e^{-\frac{ t}{\tau}}
 \cos(\Omega t) ,
 \label{uautoc}
\ee
where Greek indices denote Cartesian components and the particle index will be omitted from now.
Although, it is known that even in the non-chiral case there
are some important differences between the
two models (e.g. the stationary distribution of $f^a_\alpha$ is  circular in ABP and Gaussian in AOUP), 
previous studies \cite{fily2012athermal,farage2015effective} have shown that they share many important aspects of their
nonequilibrium behavior. It is not too unlikely to  assume that this holds true even in the
the case of angular drift.
Thus, neglecting
 autocorrelation functions of the self-propulsion ${\mathbf{f}^a_i}(t)$
beyond second order and noticing that 
 Eqs.~\eqref{eq:unitvector}  and \eqref{uautoc}  have the same functional form,
we establish the following mapping between the parameters of the two models by making the correspondence: 
\be
v_0^2= \frac{D_a}{\tau} d .
\ee

For theoretical work 
it is convenient to transform Eqs~\eqref{eq:ABPdynamics1} and 
\eqref{AOUPdyn}  into a system of equations involving the effective velocity, $\mathbf{v}$ defined as $\mathbf{v}=\mathbf{u} -\frac{\nabla \phi}{\gamma} $ and generalizing the transformation of \cite{marconi2016velocity,caprini2018active} we obtain a new set of equations (see appendix \ref{transformation}):
\bea
&&\dot{\mathbf{x}}= \mathbf{v} +\sqrt{2D_t}\boldsymbol{\xi} ,
\label{xequation}
\\
&&\dot{\mathbf{v}}= - \frac{1}{\tau} \left( \frac{1}{  \gamma}\nabla \phi +{\boldsymbol{\Gamma}}(\mathbf{x}) \cdot \mathbf{v}   \right)
+ \Omega  \hat{ \mathbf{z}} \times     \left( \mathbf{v}+ \frac{1}{  \gamma}\nabla \phi \right) 
\nonumber\\&&
+ \frac{\sqrt{2D_a}}{\tau} \boldsymbol{\eta} +\frac{{\mathbf I}-{\boldsymbol{\Gamma}}(\xx)}{\tau} \sqrt{2D_t}\boldsymbol{\xi},
\label{equazionidelmoto}
\eea
where $\mathbf{I}$ is the identity matrix and $\boldsymbol{\Gamma}$ is an effective friction tensor given by: 
\be
\Gamma_{\alpha\beta}(\mathbf{x}) = \delta_{\alpha\beta} +\frac{\tau}{\gamma}\nabla_\alpha\nabla_\beta \phi(\mathbf{x})  .
\ee
Such a transformation maps the original
problem of an active particle onto the dynamics of a fictitious passive particle, immersed in a heat bath of amplitude $\sqrt{2 D_a}/\tau$. 
The fictitious particle experiences: i) a deterministic force proportional to the potential gradient (the first term in the r.h.s
of Eq. \eqref{equazionidelmoto}), ii) a Stokes force dependent on the second derivative of the potential (the second term in the r.h.s) \cite{}, iii)  an effective Lorentz  force
 proportional to the torque $\Omega$ and orthogonal to the velocity $\mathbf{v}$ and iv)
a term proportional to $\Omega$ and orthogonal to the potential gradient.
Hereafter,
 the thermal diffusion coefficient, $D_t$ is set equal to zero, not only for simplifying the analytical work but also because it is often much smaller than $D_a$ \cite{howse2007self}. 
For the following analysis, it is convenient to write
 the Fokker-Planck equation (FPE)~\cite{risken} for the  $P(\xx,\vv,t)$ distribution corresponding to the dynamics
 \eqref{xequation} and
\eqref{equazionidelmoto}.
In the limit of vanishing thermal noise we write:
\bea
&&\frac{\partial P(\mathbf{x},\mathbf{v},t)}{\partial t}= \frac{D_a}{\tau^2} \sum_\alpha\frac{\partial^2 P(\xx,\vv,t)}{\partial v_\alpha^2}-
 \sum_\alpha v_\alpha \frac{\partial P(\mathbf{x},\mathbf{v},t)}{\partial x_\alpha}+ 
 \nonumber\\&&
 \sum_\alpha  \frac{\partial }{\partial v_\alpha}
\frac{1}{\tau}\left( \sum_\beta(\delta_{\alpha\beta}+\frac{\tau}{\gamma} \phi_{\alpha\beta}-\Omega\tau\epsilon_{\alpha\beta})v_\beta
P(\mathbf{x},\mathbf{v},t) \right) \nonumber\\&&
+ \sum_\alpha \frac{\partial }{\partial v_\alpha}
\frac{1}{\tau}\left( \sum_\beta(\delta_{\alpha\beta} - \Omega\tau\epsilon_{\alpha\beta} ) \frac{\phi_\beta}{\gamma}
P(\mathbf{x},\mathbf{v},t) \right) ,
\label{pdexv}
\eea
where for notational convenience we have adopted an explicit Cartesian representation, using 
Greek indices to denote two dimensional vector components, introduced the symbols $\phi_\alpha$ and
$\phi_{\alpha\beta}$ for the first and second partial derivatives of the potential, respectively, and
the symbol $\epsilon_{\alpha\beta} $ for the antisymmetric $2\times2$ tensor such that
$\epsilon_{yx}=-\epsilon_{xy}=1$.

 \subsection{Detailed balance}


At variance with the standard AOUP model ($\Omega=0$) 
in order to obtain the steady state solution of the FPE~\eqref{pdexv} we ought to solve
a second order partial differential equation corresponding to the vanishing of the 
phase-space divergence of the probability current~\cite{puglisi2017clausius}
${\bf I}_{\alpha}=\left((I_\alpha)_\xx,(I_\alpha)_\vv\right)$, where $\alpha=x,y$:
\be
{\bf div}_\alpha \cdot {\bf I}_\alpha=\sum_\alpha \frac{\partial}{\partial x_\alpha} (I_\alpha)_\xx+ \frac{\partial}{\partial v_\alpha} ( I_\alpha )_\vv.\ee
It is convenient to split the probability current into a reversible and an irreversible contribution
, i.e. ${\bf I}_\alpha={\bf I}^{rev}_\alpha+{\bf I}^{irr}_\alpha$, according to the their
parity under time reversal, i.e. the components $(\xx,\vv)$ of the reversible (irreversible) part of the current are transformed under time-reversal in the same (opposite) way as the time derivative of $(\xx,\vv)$, respectively.
Explicitly, the components of the probability current are:
\be
(I^{rev}_\alpha)_\xx=v_\alpha P_{st}(\mathbf{x},\mathbf{v}) ,
\label{current1}
\ee
\be
(I^{rev}_\alpha)_\vv = -\frac{1}{\tau\gamma}\left( \sum_\beta(\delta_{\alpha\beta} - \Omega\tau\epsilon_{\alpha\beta} ) \phi_\beta
\right) P_{st}(\mathbf{x},\mathbf{v}) =0 ,
\label{current2}
\ee

\be
(I^{irr}_\alpha)_\xx=0 ,
\label{current3}
\ee
\be
(I^{irr}_\alpha)_\vv =- \frac{D_a}{\tau^2} \frac{\partial P_{st}(\mathbf{x},\mathbf{v})}{\partial v_\alpha}-
\frac{1}{\tau} \sum_\beta(\delta_{\alpha\beta}+\frac{\tau}{\gamma} \phi_{\alpha\beta}-\Omega\tau\epsilon_{\alpha\beta})v_\beta
P_{st}(\mathbf{x},\mathbf{v}) .
\label{current4}
\ee
Let us remark that the $x$-component ($y$-component) of the current depends on the $y$-component 
($x$-component) of the gradient of the potential.

The system is microscopically reversible when
the detailed balance condition holds~\cite{cates2012diffusive,marconi2017heat}. Such a condition is satisfied if  ${\bf I}^{irr}_\alpha=0$ and 
${\bf div}_\alpha \cdot {\bf I}^{rev}_\alpha=0$ in the steady state.

\subsection{Unconfined chiral active motion}

In the free case  $\phi=0$,
the time-independent solution of Eq.~\eqref{pdexv} is the steady state distribution, $P_{st}$,  and  is uniform in space and has the form: 
\be
P_{st}(\xx,\vv)={\cal N}\, \exp\left(- \frac{\tau}{2 D_a}   \left (v_x^2+v_y^2\right) \right) .
\label{freechiral}
\ee
We notice that while the two-time properties of the velocity distribution depend on $\Omega$, as already seen in Eq.~\eqref{uautoc},
the presence of chirality does not affect the shape of the Maxwell-like velocity distribution \eqref{freechiral} with respect to the case $\Omega=0$, a situation which is modified when the particles are confined as we shall show below.

We, now, consider how the torque $\Omega$ modifies the diffusive properties of the particles
and find that for both ABP and AOUP the mean square displacement is given by the formula~\cite{van2008dynamics}:
\bea
&&
\langle (x_\alpha(t) -x_\alpha(0))^2\rangle=2\int_0^t dt_1 \int_0^{t_1} dt_2 \, \langle u_\alpha(t_1) u_\alpha(t_2)\rangle \nonumber\\
&&= 2 D_\Omega \Bigl( t+\frac{(1-\Omega^2\tau^2) \tau}{1+\Omega^2\tau^2} 
\left(e^{-t/\tau} \cos(\Omega t) -1 \right)
 \nonumber\\&&
 -\frac{2 \Omega\tau^2}{1+\Omega^2\tau^2} e^{-t/\tau} \sin(\Omega t) \Bigr),
 \label{diffusionlaw}
\eea
where  $D_{\Omega}=\frac{D_a}{1+\Omega^2\tau^2} $ and the thermal contribution $2 D_t t$ to the mean square displacement has been neglected.
One sees that the long-time diffusion coefficient is decreased by a factor $(1+\Omega^2\tau^2)^{-1}$
with respect to the non-chiral case \cite{ten2011brownian}, an effect which is explained by noticing that the particles
perform cycloid trajectories.


An easy check of the lack of detailed balance is provided by the free case whose distribution function is
given by \eqref{freechiral}. At variance with the AOUP with $\phi=0$, the condition that the components of the irreversible current must vanish is violated:
\bea
&&
(I^{irr}_x)_\vv = -\Omega v_y  P_{st}(\mathbf{v})  \neq 0 ,
\\
&&
(I^{irr}_y)_\vv =\Omega v_x P_{st}(\mathbf{v}) \neq 0 .
\eea

\section{Spontaneous circulation of totally confined active particles}	
\label{circular}
	
	We begin by considering a CAOUP moving in two dimensions and confined in a harmonic trap~\cite{dauchot2018dynamics}:
$$\phi (\mathbf{x})=k\frac{ (x^2+y^2)}{2}.$$
The linearity of the Ornstein-Uhlenbeck process allows to obtain the exact form of the nonequilibrium steady state,
whose
phase-space distribution reads:
\bea
&&
P(\xx,\vv)={\cal N}\, \exp\left[- \frac{\tau}{2 D_a} \Gamma  \left ((v_x+ \frac{\Gamma-1}{\Gamma}\Omega y)^2+(v_y- \frac{\Gamma-1}{\Gamma}\Omega x)^2 \right) \right]\, \nonumber\\&&
\times \exp\left[-\frac{1}{2 D_a \tau} \frac{\Gamma-1}{\Gamma}\left(\Gamma^2+\Omega^2\tau^2\right) (x^2+y^2) \right] ,
\label{circularharmonic}
\eea
where
 $\Gamma=(1+\frac{\tau}{\gamma}k)$ with $k>0$ and ${\cal N}$ is a normalization constant.

At variance with passive systems, the positional and velocity coordinates are correlated and the steady state is characterized by currents linearly increasing with $\Omega\neq 0$.
In fact, the average velocity field at fixed position $\xx$, defined as 
$$ \langle    \vv  \rangle_{ \xx}= \frac{\int d \vv   P(\xx,\vv) \vv  }{\int d \vv   P(\xx,\vv)},$$ 
is given by the formulas:
   \bea
     \langle v_x \rangle_{\xx} &=& -\frac{\Gamma-1}{\Gamma} \Omega y ,
     \\
     \langle v_y \rangle_{\xx}&=& \frac{\Gamma-1}{\Gamma} \Omega x .
     \eea
Such a field has the structure of a vortex centered at the minimum of the potential. 
The confining force pins to the origin the trajectories of different particles with the result of producing 
a single coherent macroscopic vortex, which on the contrary is absent in the free-case as 
evident from the distribution~\eqref{freechiral}.
Finally,  the velocity variance 
     \be
       \langle v_x^2 \rangle- \langle v_x\rangle^2= \langle v_y^2 \rangle- \langle v_y\rangle^2=\frac{1}{\Gamma}\frac{D_a}{\tau}
       \ee
is reduced with respect to the free case being $\Gamma>0$, but is torque independent.

We also consider
the corresponding marginal (configurational) distribution, $ 
\rho(x,y)=\int d \vv   P(\xx,\vv)$:
 \be
  \rho(x,y)={\cal N}' \exp\left(-\frac{1}{2 D_a \tau} \frac{\Gamma-1}{\Gamma}\left(\Gamma^2+\Omega^2\tau^2\right) (x^2+y^2) \right) .
  \label{marginalrot}
 \ee
We remark that the chirality determines a more concentrated distribution of particles
 near the bottom of the potential well, thus  effectively increasing its  stiffness. In other words, the chirality acts as a centripetal force
 whose strength is proportional to $ \Omega^2 \tau^2$. The non-chiral limit $\Omega \to 0$ is smoothly
 recovered and one finds  the well known AOUP distribution in the quadratic trap.
 

By inserting the exact solution \eqref{circularharmonic}
in Eqs.~\eqref{current1}-\eqref{current4}
 one finds that also in the case of harmonically confined chiral particles at variance with the non chiral AOUP
the detailed balance condition is violated due to the presence of circulating currents.
Due to this lack of detailed balance condition the unified colored noise approximation (UCNA)~\cite{hanggi1995colored}
fails even in this simple case of a chiral AOUP confined in a harmonic trap~\cite{marconi2015towards}.

\subsection{Virial Pressure}
The virial pressure formula is obtained by equating the pressure exerted by the particles on the walls
to the force per unit length that the walls exert on the particles \cite{marconi2016pressure,marini2017pressure}. Thus,
we get
\be
 2 p_v A= \langle {\mathbf \nabla} \phi \cdot \xx\rangle ,
 \ee
 where $A$ is the area and the average is performed using the steady state distribution.
 For the rotationally invariant harmonic trap we find using Eq.~\eqref{marginalrot} the following expression of the virial pressure
  $$
p_v A= \frac{D\gamma}{\Gamma+\frac{\Omega^2 \tau^2}{\Gamma}} .
$$
    Such a result shows that  the pressure decreases as $\Omega$ increases since the particles tend to 
be more localised near the bottom of the well with respect to the $\Omega=0$ case.

\section{Confinement in a channel}
\label{confinement}

Before delving into the study of the confined chiral gas,
let us briefly summarize some recent results concerning the case $\Omega=0$: 
the self-propelled particles accumulate in the proximity of repulsive walls \cite{wensink2008aggregation,elgeti2013wall}. Such a phenomenon is more evident  when
the persistence time, $\tau$, increases
 and $D_a/\tau$ is large with respect to the thermal noise amplitude, $\gamma D_t$.
 In fact, a particle pushed by the self-propulsion force along the normal to the wall remains trapped there and escapes only when the propulsion direction changes, i.e. typically after a time $\tau$. This
 mechanism leads to the formation of a peak in the proximity of each  wall and is reflected
in Eq.~\eqref{equazionidelmoto} by the presence of a 
Stokes force term, $ - \frac{1}{\tau}  \Gamma(\mathbf{x}) \cdot \mathbf{v} $, which
opposes the motion and traps the particle near the walls.
In the region where the potential gradients are negligible, instead, the particles undergo an underdamped motion
and their density is almost uniform or smoothly varying in the presence of thermal noise ($D_t\neq 0$)~\cite{caprini2018active}.


Intuitively, we expect a  
reduction of the accumulation of the particles at the walls with chirality.  In fact,
the time, $t_w$, a particle spends in front a wall is a decreasing function of $|\Omega|$, as we argue by considering  
the form of the two-time velocity autocorrelation function 
Eq.~\eqref{uautoc}, characterized by the two characteristic time-scales $\tau$ and $1/\Omega$. In the non-chiral limit,  $\Omega=0$, the residence time at the wall is 
 $t_w \approx \tau$, because after this time a particle typically inverts its self-propulsion   
and goes back to the bulk.
On the other hand, in the case $|\Omega| > 0$, the first value where the autocorrelation changes its sign is $t_{\Omega}=\pi/(2|\Omega|)$, 
the smallest zero of the cosine function in Eq.~\eqref{uautoc}.
Based upon this remark, we identify two regimes: I) $t_{\Omega} \gg \tau$, where the role of the chirality is negligible and does not affect  the distribution of the particles  in the channel, hence
$t_w \approx \tau$;
 II) $t_\Omega \leq \tau$, where the correlation~\eqref{uautoc} changes sign
 for times shorter than $\tau$. In this case, we can estimate the typical time
 needed to change orientation and leave the wall as $t_w \sim t_{\Omega}$.
Following Lee~\cite{lee2013active}, we propose a coarse-grained description where the particles belong to two different populations:
 a bulk populations of $n_b$ members and a wall population of $n_w$ elements, characterized by residence times $t_b$ and $t_w$, respectively. Due to the permanent injection of energy,  there is a continuous exchange of particles between the two populations,
 so that to achieve the steady state
we must have $n_w/t_w=n_b/t_b$. We may conclude that the wall population decreases as $t_w$ decreases, that is
when $\Omega$ grows.

In the following, we employ two different types of set-ups in order to assess quantitatively the effects of
chirality in confined systems. We begin with the study of
an infinite parabolic channel where the particles are confined only
in the $x$-direction and free to move along the $y$-direction. We shall determine the exact full steady state distribution function and show the existence of steady momentum currents induced by the chirality.

We then consider the more realistic slit case, i.e. a model where the wall and bulk regions are clearly separated and the walls are modeled by means of truncated repulsive harmonic potentials.
In this case of non-constant potential curvature, it is possible to observe the interplay
between wall accumulation and wall depletion due to the competition
between wall attraction and a chiral effective force 
pushing the particles away from the boundaries.

\subsection{Parabolic channel}
 \label{subconfinement}

\begin{figure}[!t]
\centering
\includegraphics[width=1\linewidth,keepaspectratio]{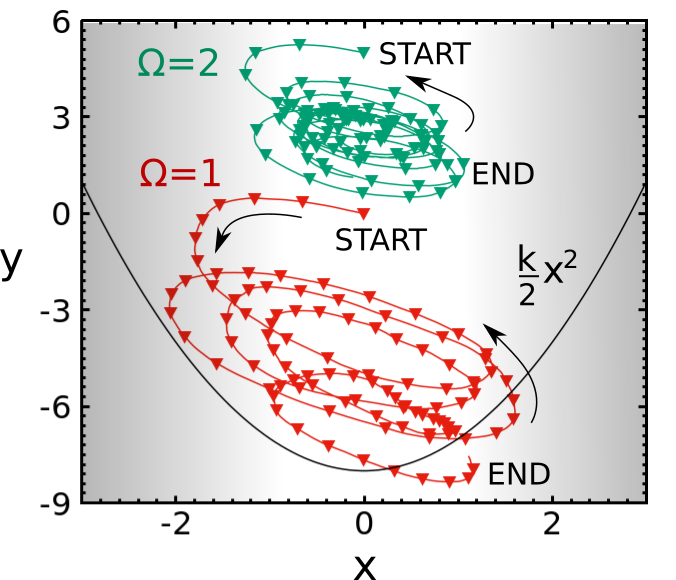}
\caption{Two typical trajectories of active chiral particles confined to a parabolic channel. The particles perform spiral motion. Notice the stronger localisation of the particle near the bottom of the well, with increasing value of the torque $\Omega$ as predicted by Eq.~\eqref{PMharmonic}. }\label{fig:trajectoryparabola}
\end{figure}

A one dimensional quadratic well $\phi (\mathbf{x}) =k\frac{ x^2}{2}$
mimics a channel limited by harmonic walls
along the $x$-direction and unbounded along the $y$-direction. 
In this case, the distribution 
depends on the two components of the velocity $\vv$, but only from the $x$-coordinate transversal to the channel. Its representation is a multivariate Gaussian distribution containing diagonal terms proportional to
$(v_x^2,v_y^2,x^2)$, but also three cross terms of type $(v_x v_y, x v_y, x v_x)$. 
The exact expression of each of the six proportionality constants, featuring in the Gaussian, is a function of the control parameters, $k,\tau,\gamma,\Omega,D_a$ and
is reported in appendix~\ref{parabolicchannel}. In the main text, to ease the presentation we write
the more transparent form of the distribution obtained by neglecting contributions
to the coefficients beyond the linear order in $\Omega$:
\bea
&&
P(\xx,\vv)\approx {\cal N}\, \exp\left[- \frac{\tau}{2 D_a}  \left (\Gamma  v_x^2+ v_y^2
+2\frac{\Gamma-1}{\Gamma}  \Omega\tau v_x v_y  \right) \right]\, \nonumber\\
&&
 \exp\left[-\frac{\Gamma-1}{2 D_a \tau} \Gamma x^2  \right] \,
  \exp\left[\Omega \frac{\tau}{D_a}  \frac{\Gamma-1}{\Gamma} v_y x \right] .
\eea
By integrating over $v_x$ and $v_y$ we obtain
the $x$-dependent average velocity along the $y$-direction as a function of $\Omega$
         \be  
  \langle v_y \rangle_x = \frac{\Gamma-1}{\Gamma} \Omega x .
  \label{vyave}
  \ee
  The result \eqref{vyave} is valid to all orders in $\Omega$ and for
  $\Omega>0$ predicts the existence of a current parallel to the walls and directed along the downward $y$-direction for $x<0$ and upwards for $x>0$, while reversing the sign of 
  $\Omega$  the current in the two halves of the channel changes sign.  
Instead, there is no net average current ($\langle v_x \rangle_x $ along the $x$-direction. Finally,  
 using the exact form of the distribution given by Eq.~\eqref{exactpxv} we obtain
 associated marginalized positional distribution function
  \be
  \rho(x)={\cal N}\exp\left(-\frac{1}{2 D_a \tau} \frac{\Gamma-1}{\Gamma}\left(\Gamma^2+\Omega^2\tau^2\right) x^2\right)
  \label{PMharmonic}
  \ee
 which shows that the density near the center
  is enhanced with respect to the $\Omega=0$ case as if an extra effective
  potential $\frac{\Gamma-1}{\Gamma}\Omega^2\tau^2\ x^2/2$ was
  pushing the particles towards the midpoint.
  In Fig.~ \ref{fig:trajectoryparabola} we display two typical trajectories for two different values of $\Omega$,
  which become more and more localised as the chirality increases.
 
Unlike the rotationally invariant case, the variances of the velocity of the particles weakly depend on the value of $\Omega$ as shown by the exact solution of appendix \ref{parabolicchannel}:
 $$\langle v_x^2 \rangle_{x=0}  =\frac{D_a}{\tau} \,  \frac{\Omega^2\tau^2+\Gamma}{ \Omega^2\tau^2+\Gamma^2 } ,$$
 $$
   \langle v_y^2 \rangle_{x=0}  =\frac{D_a}{\tau \Gamma} \,  \frac{\Omega^2\tau^2+\Gamma^3}{ \Omega^2\tau^2+\Gamma^2 }
$$
and 
\be
  \langle v_x v_y \rangle_{x=0} =-\frac{D_a}{\tau}(\Gamma-1) \frac{\Omega\tau}{ \Omega^2\tau^2+\Gamma^2 } ,
 \ee
where the subscript means that the average refers to the midpoint $x=0$.

\subsection{The slit }

We turn, now, to study a different
type of channel where
the stiffness of the walls and the width of the channel can be varied independently.
A two-dimensional collection of independent chiral active particles is confined between two parallel
repulsive soft walls, exerting a force  piece-wise linear, characterized by an elastic constant $k$ according to the formula:
\be
F_w(x)=  k(x+L) \Theta(-x-L) - k (x-L)\Theta(x-L) ,
\label{wallforce}
\ee
where $\Theta$ is the Heaviside step-distribution.
We choose a large value of $k$ so that the penetration inside the wall is negligible. 
 The space between the walls extending from $x=-L $ to $x=L$, instead, forms a force-free region.
 The setup, recently studied in the case of non-chiral particles, $\Omega=0$, provides a more realistic
description of a straight capillary because 
we may clearly distinguish a boundary, potential region from a bulk-like region~\cite{caprini2018active}.
In the numerical simulations, in order to model an infinite vertical channel
we assume periodic boundary conditions along the  $y$-direction. At variance with the non chiral case, where the problem becomes effectively one-dimensional, in the $\Omega\neq 0$ case, as already found in the previous subsection \ref{subconfinement}, both  $x$ and $y$ directions matter and the system develops a steady vertical momentum current and the bulk region becomes overpopulated with respect to the $\Omega=0$ case.

\begin{figure}[!t]
\centering
\includegraphics[width=1\linewidth,keepaspectratio]{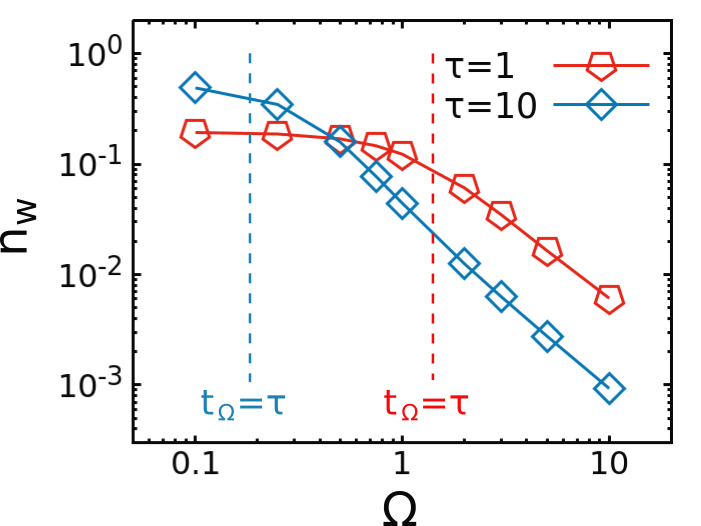}
\caption{Density at the walls $n_w$  versus $\Omega$ for two different values of $\tau=1$ (red pentagons) and $10$ (blue diamonds). The parameters are: $D_a=\tau$, $\gamma=1$, $k=10$, $L=8$, and the walls are at positions  $x=\pm L$.}\label{fig:accumulation}
\end{figure}

 In Fig.~\ref{fig:accumulation}, we plot the density at the walls, $n_w$,  as a function of $\Omega$, for two different parameter configurations: we fix the ratio $D_a /\tau = 1$ and perform numerical simulations for $\tau= 1$ and $\tau = 10$. We evaluate $n_w$ by counting the number of particles in the regions $x\leq -L$ and $x \geq L$, for the left and right wall, respectively. 
 In the regime I) identified in section \ref{confinement} the decrease of $n_w$ is slower than $1/|\Omega|$, since the effect of the chirality  does not affect the wall population being $t_{\Omega}>\tau$. 
On the other hand, in regime II) the scaling $1/\Omega$ fairly agrees with the data
for both choices of $\tau$.

\begin{figure}[!t]
\centering
\includegraphics[width=1\linewidth,keepaspectratio]{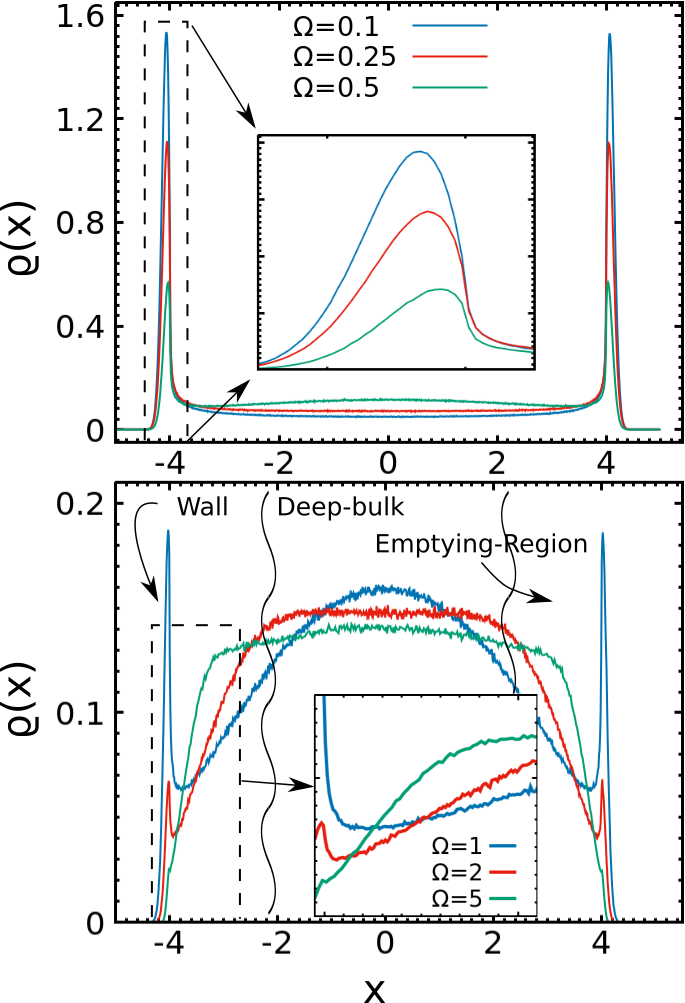}
\caption{Density profile $\rho(x)$ for different values of $\Omega$ as shown in the legend: $\Omega=0.1, 0.25, 0.5$ (Panel (a)), $\Omega=1, 2, 5$ (Panel (b)). The control parameters are: $D_a=10$, $\tau=10$, $\gamma=1$, $k=10$ and $L=4$.
In the insets, we report the zoomed profiles in regions delimited by dashed vertical lines.
}\label{fig:PX}
\end{figure}

In Fig.\ref{fig:PX} we show the marginalized distribution $\rho(x)$ for a system of active chiral particles
subject to the dynamics of 
Eqs.~\eqref{xequation}-\eqref{equazionidelmoto} in the presence of the wall
force \eqref{wallforce} . We consider the dependence of the profile on the internal torque by varying  $\Omega$ and keeping 
$D_a$ and $\tau$ fixed to values such that
there is an appreciable accumulation at the walls.
Upon introducing a small chirality the accumulation at the walls is reduced, as shown in Fig.\ref{fig:PX} (a).
In an intermediate regime of $\Omega$, the role of the chirality is more consistent: on one hand, particles accumulate in front of the wall (as in the case $\Omega=0$), on the other hand, the profile in the bulk is no longer flat-like. In particular, we observe an emptying phenomenon in a layer near the wall favoring the accumulation in the inner region, as shown in Fig.\ref{fig:PX} (b).
A further increase of $\Omega$ depletes the density near the walls and enhances it in the bulk region until the wall region remains almost completely empty. For these values of $\Omega$, the situation is completely inverted with respect to the case $\Omega=0$ and the wall behaves as if the effective wall-potential were repulsive.\\

We stress that, at variance with the harmonically confined CAOUP of Sec.~\ref{subconfinement}, it is possible to observe a transition from a situation where particles accumulate near the walls to a situation where they accumulate in the bulk.



We, now, explain the process which leads to the emptying of the wall region and to the enrichment of the
central region in the case of the channel when $\Omega\neq 0$
and to this purpose we rewrite  Eq.~\eqref{equazionidelmoto}  under the explicit form:
\bea
&&\dot{v}_x = - \Bigl[1 +(\Gamma-1) \left(\Theta(-x-L) +\Theta(x-L)\right) \Bigr]\frac{v_x}{\tau} - \Omega v_y 
\nonumber\\
&&-\Theta(-x-L) \frac{k}{\gamma \tau}(x+L) -\Theta(x-L) \frac{k}{\gamma \tau}(x-L) +  \frac{\sqrt{2D_a}}{\tau}\eta_x \nonumber\\
&&\dot{v}_y = - \frac{v_y}{\tau} + \Omega v_x + \frac{\Omega}{\gamma \tau}\left[ k(x+L) \Theta(-x-L) + k(x-L)\Theta(x-L)\right] \nonumber\\
&& + \frac{\sqrt{2D_a}}{\tau}\eta_y \nonumber .
\eea

\begin{figure}[!t]
\centering
\includegraphics[width=1\linewidth,keepaspectratio]{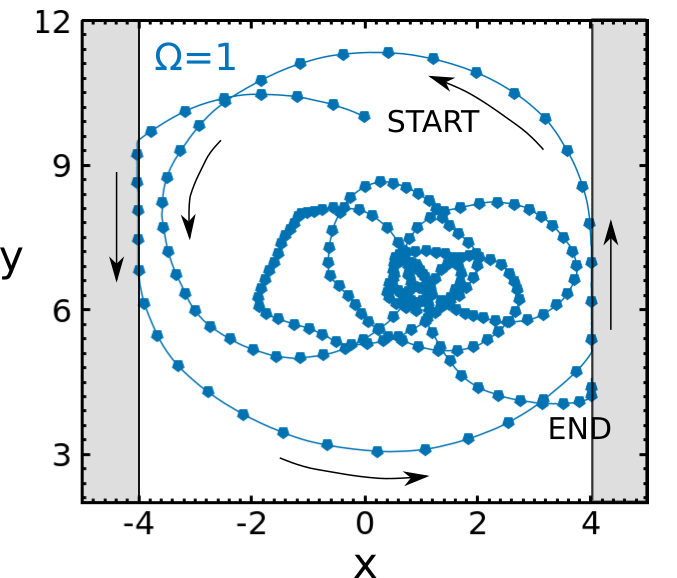}
\caption{A typical trajectory of an active chiral particle confined to a slit-like channel.
In the inner region, the particle performs spiral-like motion, while at the walls it slides vertically. Notice that the sliding is downwards at the left wall and upwards at the right wall.}\label{fig:traiettoriaslit}
\end{figure}

One sees that on one side
the chirality acts as an effective magnetic field of strength $\Omega$  
periodically rotating the velocity direction without changing its magnitude and the other side, for $|x|>L$, in combination with the 
wall repulsion it produces a tangential force always along the $y$ direction.
As a result, the particle accelerates and gains  "kinetic energy".
Such a force, proportional to $k\Omega/\tau$ only acts in the boundary region and generates a vertical momentum current having opposite directions at the two walls.
When a particle penetrates the potential region, its velocity components $v_x$ is strongly damped by the large Stokes force $-\Gamma v_x/\tau$ and thus it remains trapped there. At the same time,  the $ v_y$ component of the velocity rapidly increases under the action of the tangential force. This process continues for a time $t_\Omega$
after which the torque rotates the velocity vector and transfers the accumulated "kinetic energy" to the $x$ component of the velocity pointing towards the bulk.
 In the regime of large $|\Omega|\tau\gg 1$, the tangential field acts on a shorter time scale than the scale of the dissipative Stokes force
 and the velocity is nearly unaffected by the friction.
The particle leaves the wall  with a velocity proportional to $\Omega$ and enters the potential-free region
where it performs a spiral motion with a large initial radius,
due to the energy accumulated at the wall.
This radius continuously shrinks due to the dissipation caused by the Stokes bulk force,
as shown in Figure \ref{fig:traiettoriaslit}.
After a time $\sim \tau$ particles  the dissipation becomes relevant and $\langle|v| \rangle\sim \sqrt{D_a /\tau}$,
i.e. the steady state typical velocity. 
The process here described goes on forever since a fluctuation of the self-propulsion can drive a bulk particle to reach again the wall region to start a new cycle.
It is clear, that the removal of particles from the walls due to $\Omega$ has the effect of depleting the wall accumulation and increasing the bulk population.

To test the existence of large surface currents leading to the formation of an intermediate emptying region we have studied the profile of the vertical velocity.
In the two panels Fig.~\ref{fig:moment} we display the average velocity $\langle v_y\rangle_x$ as a function of the
distance from the center of the slit $x$ for two different values of $\Omega$. We notice the monotonic behavior of the velocity, which can be described by a hyperbolic sine function since the average velocity in the left-hand side has the opposite direction with respect to velocity in the right-hand side.
As discussed above, the mechanism which causes such a behavior is 
 the interplay between the wall force and the chiral force,
however, in this case, the non-uniformity of the Stokes force in the $x$-direction leads to a richer structure
 both in the density and in the velocity profiles than the one observed in the parabolic channel. In fact, the Stokes force opposing the motion along the $x$ direction is large only in the
wall region $|x|>L$, while the centripetal force $\Omega  \hat{ \mathbf{z}} \times   \mathbf{v}$ is spatially uniform.

\subsection{Momentum profile in the potential-free region $|x|<L$ and effective viscosity}

\begin{figure}[!b]
\centering
\includegraphics[width=1.\linewidth,keepaspectratio]{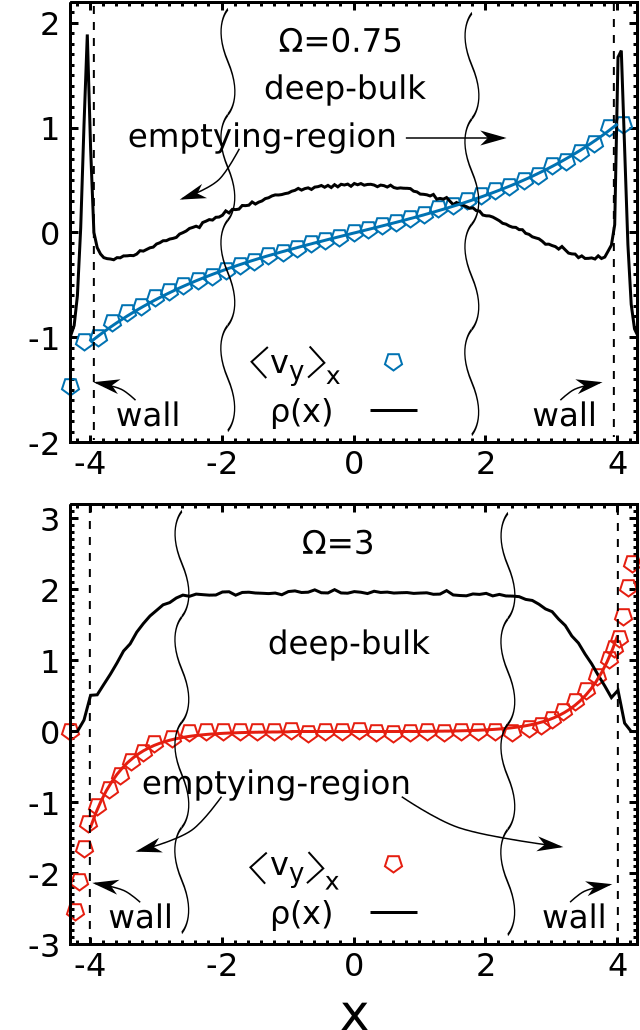}
\caption{Velocity and density profiles for the slit-like channel.
Left panel: case  $\Omega=0.75$. Mean vertical velocity $\langle v_y\rangle_x$ as a function of the $x$-coordinate (blue symbols) and theoretical prediction (blue line) Eq.~\eqref{jyvsx}. The black line indicates
the corresponding density profile $\rho(x)$.
Right panel: case  $\Omega=3.$. Mean vertical velocity $\langle v_y\rangle_x$ versus $x$ (red symbols) and theoretical prediction (red line)  Eq.~\eqref{jyvsx}. The black line indicates
the density profile $\rho(x)$.
For both panels the control parameters are: $\tau=10$, $\gamma=1$, $D_a=10$, $L=4$.}\label{fig:moment}
\end{figure}

As shown in the two panels of Fig.~\ref{fig:moment}, the vertical component of the average velocity
field varies in the region bounded by the walls. Hereafter, we derive the equations for the
current by projecting the Fokker-Planck equation onto an appropriate hydrodynamic space.
By integrating the FPE~\eqref{pdexv} over $v_x,v_y$ we obtain the continuity equation:
\be
\frac{\partial \rho(\xx,t)}{\partial t}+\frac{\partial j_x(\xx,t)}{\partial x}+\frac{\partial j_y(\xx,t)}{\partial y}=0
\ee
with 
$$j_\alpha(\xx,t)=\int dv_x dv_y  v_\alpha P(\xx,\vv,t) .
$$
In the steady state $\frac{\partial j_x(x,t)}{\partial x}=0$ and
by symmetry the particle current
 $ {\bf j}$ is a function of $x$ only. From the continuity equation and the wall boundary conditions it follows that $j_x(x)=0$, while the vertical component
 $j_y(x)$ varies with $x$.
Next, we define the pressure tensor ${\mathcal P}$ as: 
 $${\mathcal P}_{\alpha\beta}(\xx)=\int dv_x dv_y  v_\alpha v_\beta P(\xx,\vv) ,
$$
multiply the stationary FPE by $v_y$ and $v_x$, respectively, and integrate over the velocity.
In the potential-free region, $|x|<L$,
we obtain the following equations relating the components of the pressure tensor to the current $j_y$ :
 \be
- \frac{1}{\tau} j_y(x)
-\frac{\partial {\mathcal P}_{xy}(x)}{\partial x} =0 ,
\label{jy2}
\ee
\be
-\Omega j_y(x)
-\frac{\partial {\mathcal P}_{xx}(x)}{\partial x} =0 .
\label{jy1}
\ee
 In order to determine $j_y(x)$ from Eq.~\eqref{jy2}, we must relate it to $ {\mathcal P}_{xy}(x)$, a task which will be pursued hereafter using a simple kinetic argument.

In the situation depicted in Fig.~\ref{fig:moment}, let us consider the 
average vertical momentum flux $J_y^{+}$ per unit time and per unit length crossing a segment at $x=x_0$
(with $-L<x_0<L$) and originating from the region $x<x_0$.
 We assume that: a) the particles
 passing from the one side to the other of the unit vertical segment at $x_0$ 
 move at the constant average velocity,  $ |\overline v_x|$ b) the physical space
 can be divided into cells of linear size  $\lambda_\Omega$, the smallest distance over which the mean values of the physical observables vary. Under these hypotheses, the magnitude of  $J_y^{+}$  can be written
as:
\be
J_y^{+}(x_0)=\frac{1}{4}  |\overline v_x| j_y(x_0-\lambda_\Omega) .
\ee
Similarly,
the momentum flux of the particles coming from the right is
\be
J_y^{-}(x_0)=-\frac{1}{4}  |\overline v_x| j_y(x_0+\lambda_\Omega) ,
\ee
where
the geometrical factor $1/4$ takes into account the different possible directions of the velocities of the particles
under the assumption of isotropy of the velocity distribution.
The result is a net transfer of the $y$-component of momentum across the segment $x=x_0$ given by:
\be
J_y^{+}(x_0)-J_y^{-}(x_0)\approx-\frac{1}{2}  \lambda_\Omega
  |\overline v_x|  \frac{\partial j_y}{\partial x}|_{x_0} .
 \ee
 The rate of change of momentum per unit length, $(J_y^{+}-J_y^{-})$, due to the diffusion of the particles in the horizontal direction 
 is equivalent to the shear force per unit length exerted by the particles with $x<x_0$ on the particles with $x>x_0$, that is to the off-diagonal component of the pressure tensor:
\be
{\mathcal P}_{xy}=J_y^{+}-J_y^{-}= -\nu_\Omega \frac{\partial j_y}{\partial x}
 \ee
 with $\nu_\Omega=\frac{1}{2}  \lambda_\Omega
  | \overline v_x|  $.
 The same kind of elementary kinetic argument, this time replacing the transported quantity $j_y(x)$ by $\rho(x)$, predicts that a  gas of tracer chiral active particles of non uniform density $\rho(x)$
 generates a net flux $J_N$, defined as the number of particles  per unit time crossing a unit vertical segment  
 according to the formula:
 \be
 J_N=-\frac{1}{2}  \lambda_\Omega
  |\overline v_x|  \frac{\partial \rho(x)}{\partial x}|_{x_0} .
 \ee
The ratio between  $-J_N$ and $- \frac{\partial \rho(x)}{\partial x}$ defines the 
the self-diffusion coefficient of chiral particles
$D_\Omega=\frac{1}{2}  \lambda_\Omega   |\overline v_x|$. One concludes that
the kinematic shear viscosity and the diffusion coefficient are equal: $\nu_\Omega=D_\Omega$, a result well known in kinetic theory of dilute gases.
By comparing this result with the exact calculation of the long-time diffusion coefficient $D_{\Omega}=\frac{D_a}{1+\Omega^2\tau^2} $ given by
Eq.~\eqref{diffusionlaw}
we can now directly determine the value of $\nu_\Omega$ and thus obtain
\be
{\mathcal P}_{xy}=-  \frac{D_a}{1+\Omega^2\tau^2} \frac{\partial j_y}{\partial x} .
\ee
Finally, substituting in Eq.~\eqref{jy2} we obtain 
\be
 \frac{\partial^2 j_{y}}{\partial x^2} =\frac{1+\Omega^2\tau^2}{D_a \tau} j_y
 \ee 
 whose solution is
 \be
j_y(x)=A \sinh(\frac{x}{\sqrt{\nu_\Omega\tau}})=A \sinh \left(\frac{\sqrt{1+\Omega^2\tau^2 }}{\sqrt{D_a\tau}}x \right) ,
\label{jyvsx}
\ee
where the constant $A$ is fixed by the odd boundary conditions at $x=\pm L$.
Using Eq.~\eqref{jy1} we can obtain ${\mathcal P}_{xx}$:
\be
{\mathcal P}_{xx}(x)=\int_{-L}^x dx' j_y(x')={\mathcal P}_0 - A \Omega \sqrt{\nu_\Omega\tau}  \cosh(\frac{x}{\sqrt{\nu_\Omega\tau}}) ,
\ee
where  the constant ${\mathcal P}_0$ is determined by the boundary conditions.
Let us remark that $j_y(x)$ is proportional to  $\langle v_y\rangle_x$ in the regime where the density $\rho(x)$ varies slower than $\langle v_y\rangle_x$ itself. Under this condition 
it is legitimate to compare directly the solution $j_y(x)$ of Eq.~\eqref{jyvsx} with
the profile  $\langle v_y\rangle_x$ of  Fig.~\ref{fig:moment}.
The present estimate of the shear kinematic viscosity is of the right order and we find
a quantitative agreement between the prediction of of Eq.~\eqref{jyvsx} 
and the numerical simulation by a simple rescaling of
the characteristic length: $\sqrt{\nu_\Omega \tau} \to \sqrt{2 \nu_\Omega\tau }$.

\section{Conclusions}
\label{conclusions}
In this paper, we have introduced an extension of the active Ornstein-Uhlenbeck model
to study the behavior of a gas of active chiral particles under confinement. The handedness
of particles' motion has been accounted for by means of an effective torque term
in the self-propulsion forcing. The presence of the torque has a profound influence on the properties
of the AOUP as it breaks the detailed balance condition even in the unconfined case.
Under confinement, we find that the torque is responsible for the appearance of steady momentum currents
and of reduction of the accumulation of the particles at the container's boundaries.
Explicit illustrations of these phenomena are shown by means 
of the exact solution of the stationary Fokker-Planck equation in two cases of parabolic confinement.
In a more realistic case of confinement by stiff walls,
beside the emptying of the boundary region and the enrichment of the bulk region,
we find a structure the velocity field akin to the one observed
in sheared viscous fluids. This behavior is explained in terms of a simple kinetic argument
and the velocity profile is reproduced.

Since the linear propulsive and rotational behavior can independently be tuned, by changing $\tau$ and $\Omega$ respectively, one could drive more efficiently the motion of active particles for instance by partly suppressing their diffusivity, or obtain more efficient harmonic traps. 
Active rotation does also have an impact on the properties of interacting systems: Liao and Klapp have found that it
generally opposes motility-induced clustering and phase separation, as demonstrated by a narrowing of the coexistence region
of two-dimensional chiral ABP upon increase of the propulsion angular velocity~\cite{liao2018clustering}. A similar behaviour should also be observablein systems of interacting CAOUP.

Finally, concerning the practical interest of the systems studied, one could envisage the possibility of selecting chiral active particles on the basis of their handedness in pharmaceutical or biotechnological applications where chiral levogyre or dextrogyre properties correspond to different functionalities~\cite{mijalkov2013sorting,meinhardt2012separation}.

\section*{Conflicts of interest}
There are no conflicts of interest to declare.

\section*{Acknowledgements}

We thank Andrea Puglisi and Claudio Maggi for illuminating discussions.

\appendix

\section{Transformation of variables}
\label{transformation}
In this appendix, we derive the governing equations for the transformed variables $(x_i,v_i)$. Let us consider
the AOUP equation:
\be
\dot  x _i= -\frac{1}{\gamma}\frac{\partial \phi}{\partial x_i} + \sqrt{2D_t}\xi_i + u_i 
\label{xpunto}
\ee
and define the "velocity", $v_i$:
$v_i=\dot x_i-\sqrt{2 D_t} \xi_i(t)$.  We differentiate \eqref{xpunto} 
and eliminate $u_i$ with the help of Eq.~\eqref{AOUPdyn} and use again \eqref{xpunto}.
Adopting Stratonovich convention for the derivatives of $\phi$ we finally get:
\bea
\dot x_i&=&v_i +\sqrt{2 D_t}  \xi_i
\label{sdotxi}
\\
\dot v_i&=& -\frac{1}{\tau}[ \frac{1}{\gamma} \frac{\partial \phi}{\partial x_i} +v_i] 
+\Omega  \epsilon_{ik}[ v_k+\frac{1}{\gamma} \frac{\partial \phi}{\partial x_k}]
 -\frac{1}{ \gamma}  \frac{\partial^2 \phi}{\partial x_i \partial x_k} v_k
 \nonumber\\
&&
+\frac{ \sqrt{2 D_a}}{ \tau} \eta_i- \frac{\sqrt{2 D_t}}{\gamma}  \frac{\partial^2 \phi}{\partial x_i \partial x_k}\xi_k
\label{dotvi}
\eea
where we used the Einstein repeated indexes convention.

\section{Exact distribution in the parabolic channel}
\label{parabolicchannel}
Hereafter, we report the exact phase-space distribution in the case of a parabolic channel. Its form as discussed in the main text depends on six constants $A,K_1,K_2,K_3,M_1,N_2$ according to the formula:
\bea
&&
P(\xx,\vv)= {\cal N}\,  \exp\left(-\frac{A}{2 } x^2  \right) \, \exp\left(- \frac{K_1}{2 }   v_x^2 - \frac{K_2}{2 } v_y^2  -K_3  v_x v_y  \right)\nonumber\\
&&
 \times \exp\left(-  M_1  v_x x -    N_2 v_y x \right) .  
\label{exactpxv}
\eea
For the sake of notational simplicity we define:
$$\Delta=\Omega^2\tau^2 + \Gamma^2$$
and write the following six equations for the coefficients in terms of the independent control parameters:
$$
A=
\frac{k}{D_a \gamma }\frac{ \Omega^4\tau^4 +2 \Omega^2\tau^2 + 3\Omega^2\tau^2\left( \Gamma -1  \right) + \Gamma^3  }{\Delta} ,
$$
$$
K_1
=\frac{\tau}{D_a} \frac{\Omega^2\tau^2+\Gamma^3}{\Delta} ,
$$
$$
K_2=
\frac{\tau}{D_a}\Gamma \frac{\Omega^2\tau^2+\Gamma}{\Delta} ,
$$
$$
K_3
=\frac{\Omega  \tau^3}{D_a \gamma}  \frac{ \Gamma} { \Delta}  k ,
$$
$$
M_1
=-\Omega^2\tau^4 \frac{1}{D_a \gamma} \frac{k^2}{\gamma} \frac{ 1} { \Delta} ,
$$
$$
N_2=
-\frac{\Omega  \tau^2}{D_a \gamma} \frac{\Omega^2\tau^2+\Gamma} {\Delta} k .
$$

The average velocity along the y is
         \be  
  \langle v_y \rangle_x =  \frac{\Gamma-1}{\Gamma} \Omega x
  \ee
  and is proportional to $\frac{\Omega \tau}{\gamma}k  $.

\bibliographystyle{rsc} 

\bibliography{caoup.bib} 

\providecommand*{\mcitethebibliography}{\thebibliography}
\csname @ifundefined\endcsname{endmcitethebibliography}
{\let\endmcitethebibliography\endthebibliography}{}
\begin{mcitethebibliography}{45}
\providecommand*{\natexlab}[1]{#1}
\providecommand*{\mciteSetBstSublistMode}[1]{}
\providecommand*{\mciteSetBstMaxWidthForm}[2]{}
\providecommand*{\mciteBstWouldAddEndPuncttrue}
  {\def\EndOfBibitem{\unskip.}}
\providecommand*{\mciteBstWouldAddEndPunctfalse}
  {\let\EndOfBibitem\relax}
\providecommand*{\mciteSetBstMidEndSepPunct}[3]{}
\providecommand*{\mciteSetBstSublistLabelBeginEnd}[3]{}
\providecommand*{\EndOfBibitem}{}
\mciteSetBstSublistMode{f}
\mciteSetBstMaxWidthForm{subitem}
{(\emph{\alph{mcitesubitemcount}})}
\mciteSetBstSublistLabelBeginEnd{\mcitemaxwidthsubitemform\space}
{\relax}{\relax}

\bibitem[Ramaswamy(2010)]{ramaswamy2010mechanics}
S.~Ramaswamy, \emph{The Mechanics and Statistics of Active Matter}, 2010,
  \textbf{1}, 323--345\relax
\mciteBstWouldAddEndPuncttrue
\mciteSetBstMidEndSepPunct{\mcitedefaultmidpunct}
{\mcitedefaultendpunct}{\mcitedefaultseppunct}\relax
\EndOfBibitem
\bibitem[Bechinger \emph{et~al.}(2016)Bechinger, Di~Leonardo, L{\"o}wen,
  Reichhardt, Volpe, and Volpe]{bechinger2016active}
C.~Bechinger, R.~Di~Leonardo, H.~L{\"o}wen, C.~Reichhardt, G.~Volpe and
  G.~Volpe, \emph{Reviews of Modern Physics}, 2016, \textbf{88}, 045006\relax
\mciteBstWouldAddEndPuncttrue
\mciteSetBstMidEndSepPunct{\mcitedefaultmidpunct}
{\mcitedefaultendpunct}{\mcitedefaultseppunct}\relax
\EndOfBibitem
\bibitem[Marchetti \emph{et~al.}(2013)Marchetti, Joanny, Ramaswamy, Liverpool,
  Prost, Rao, and Simha]{marchetti2013hydrodynamics}
M.~Marchetti, J.~Joanny, S.~Ramaswamy, T.~Liverpool, J.~Prost, M.~Rao and R.~A.
  Simha, \emph{Reviews of Modern Physics}, 2013, \textbf{85}, 1143\relax
\mciteBstWouldAddEndPuncttrue
\mciteSetBstMidEndSepPunct{\mcitedefaultmidpunct}
{\mcitedefaultendpunct}{\mcitedefaultseppunct}\relax
\EndOfBibitem
\bibitem[Romanczuk \emph{et~al.}(2012)Romanczuk, B{\"a}r, Ebeling, Lindner, and
  Schimansky-Geier]{romanczuk2012active}
P.~Romanczuk, M.~B{\"a}r, W.~Ebeling, B.~Lindner and L.~Schimansky-Geier,
  \emph{The European Physical Journal Special Topics}, 2012, \textbf{202},
  1--162\relax
\mciteBstWouldAddEndPuncttrue
\mciteSetBstMidEndSepPunct{\mcitedefaultmidpunct}
{\mcitedefaultendpunct}{\mcitedefaultseppunct}\relax
\EndOfBibitem
\bibitem[Lauga \emph{et~al.}(2006)Lauga, DiLuzio, Whitesides, and
  Stone]{lauga2006swimming}
E.~Lauga, W.~R. DiLuzio, G.~M. Whitesides and H.~A. Stone, \emph{Biophysical
  journal}, 2006, \textbf{90}, 400--412\relax
\mciteBstWouldAddEndPuncttrue
\mciteSetBstMidEndSepPunct{\mcitedefaultmidpunct}
{\mcitedefaultendpunct}{\mcitedefaultseppunct}\relax
\EndOfBibitem
\bibitem[Friedrich and J{\"u}licher(2008)]{friedrich2008stochastic}
B.~Friedrich and F.~J{\"u}licher, \emph{New Journal of Physics}, 2008,
  \textbf{10}, 123025\relax
\mciteBstWouldAddEndPuncttrue
\mciteSetBstMidEndSepPunct{\mcitedefaultmidpunct}
{\mcitedefaultendpunct}{\mcitedefaultseppunct}\relax
\EndOfBibitem
\bibitem[DiLuzio \emph{et~al.}(2005)DiLuzio, Turner, Mayer, Garstecki, Weibel,
  Berg, and Whitesides]{diluzio2005escherichia}
W.~R. DiLuzio, L.~Turner, M.~Mayer, P.~Garstecki, D.~B. Weibel, H.~C. Berg and
  G.~M. Whitesides, \emph{Nature}, 2005, \textbf{435}, 1271\relax
\mciteBstWouldAddEndPuncttrue
\mciteSetBstMidEndSepPunct{\mcitedefaultmidpunct}
{\mcitedefaultendpunct}{\mcitedefaultseppunct}\relax
\EndOfBibitem
\bibitem[Loose and Mitchison(2014)]{loose2014bacterial}
M.~Loose and T.~J. Mitchison, \emph{Nature cell biology}, 2014, \textbf{16},
  38\relax
\mciteBstWouldAddEndPuncttrue
\mciteSetBstMidEndSepPunct{\mcitedefaultmidpunct}
{\mcitedefaultendpunct}{\mcitedefaultseppunct}\relax
\EndOfBibitem
\bibitem[K{\"u}mmel \emph{et~al.}(2013)K{\"u}mmel, ten Hagen, Wittkowski,
  Buttinoni, Eichhorn, Volpe, L{\"o}wen, and Bechinger]{kummel2013circular}
F.~K{\"u}mmel, B.~ten Hagen, R.~Wittkowski, I.~Buttinoni, R.~Eichhorn,
  G.~Volpe, H.~L{\"o}wen and C.~Bechinger, \emph{Physical review letters},
  2013, \textbf{110}, 198302\relax
\mciteBstWouldAddEndPuncttrue
\mciteSetBstMidEndSepPunct{\mcitedefaultmidpunct}
{\mcitedefaultendpunct}{\mcitedefaultseppunct}\relax
\EndOfBibitem
\bibitem[L{\"o}wen(2016)]{lowen2016chirality}
H.~L{\"o}wen, \emph{The European Physical Journal Special Topics}, 2016,
  \textbf{225}, 2319--2331\relax
\mciteBstWouldAddEndPuncttrue
\mciteSetBstMidEndSepPunct{\mcitedefaultmidpunct}
{\mcitedefaultendpunct}{\mcitedefaultseppunct}\relax
\EndOfBibitem
\bibitem[Blakemore(1975)]{blakemore1975magnetotactic}
R.~Blakemore, \emph{Science}, 1975, \textbf{190}, 377--379\relax
\mciteBstWouldAddEndPuncttrue
\mciteSetBstMidEndSepPunct{\mcitedefaultmidpunct}
{\mcitedefaultendpunct}{\mcitedefaultseppunct}\relax
\EndOfBibitem
\bibitem[Lef{\`e}vre and Bazylinski(2013)]{lefevre2013ecology}
C.~T. Lef{\`e}vre and D.~A. Bazylinski, \emph{Microbiology and Molecular
  Biology Reviews}, 2013, \textbf{77}, 497--526\relax
\mciteBstWouldAddEndPuncttrue
\mciteSetBstMidEndSepPunct{\mcitedefaultmidpunct}
{\mcitedefaultendpunct}{\mcitedefaultseppunct}\relax
\EndOfBibitem
\bibitem[Ten~Hagen \emph{et~al.}(2015)Ten~Hagen, Wittkowski, Takagi,
  K{\"u}mmel, Bechinger, and L{\"o}wen]{ten2015can}
B.~Ten~Hagen, R.~Wittkowski, D.~Takagi, F.~K{\"u}mmel, C.~Bechinger and
  H.~L{\"o}wen, \emph{Journal of Physics: Condensed Matter}, 2015, \textbf{27},
  194110\relax
\mciteBstWouldAddEndPuncttrue
\mciteSetBstMidEndSepPunct{\mcitedefaultmidpunct}
{\mcitedefaultendpunct}{\mcitedefaultseppunct}\relax
\EndOfBibitem
\bibitem[Najafi and Golestanian(2004)]{najafi2004simple}
A.~Najafi and R.~Golestanian, \emph{Physical Review E}, 2004, \textbf{69},
  062901\relax
\mciteBstWouldAddEndPuncttrue
\mciteSetBstMidEndSepPunct{\mcitedefaultmidpunct}
{\mcitedefaultendpunct}{\mcitedefaultseppunct}\relax
\EndOfBibitem
\bibitem[van Teeffelen and L{\"o}wen(2008)]{van2008dynamics}
S.~van Teeffelen and H.~L{\"o}wen, \emph{Physical Review E}, 2008, \textbf{78},
  020101\relax
\mciteBstWouldAddEndPuncttrue
\mciteSetBstMidEndSepPunct{\mcitedefaultmidpunct}
{\mcitedefaultendpunct}{\mcitedefaultseppunct}\relax
\EndOfBibitem
\bibitem[Ebeling \emph{et~al.}(1999)Ebeling, Schweitzer, and
  Tilch]{ebeling1999active}
W.~Ebeling, F.~Schweitzer and B.~Tilch, \emph{BioSystems}, 1999, \textbf{49},
  17--29\relax
\mciteBstWouldAddEndPuncttrue
\mciteSetBstMidEndSepPunct{\mcitedefaultmidpunct}
{\mcitedefaultendpunct}{\mcitedefaultseppunct}\relax
\EndOfBibitem
\bibitem[Cates and Tailleur(2013)]{cates2013active}
M.~Cates and J.~Tailleur, \emph{EPL (Europhysics Letters)}, 2013, \textbf{101},
  20010\relax
\mciteBstWouldAddEndPuncttrue
\mciteSetBstMidEndSepPunct{\mcitedefaultmidpunct}
{\mcitedefaultendpunct}{\mcitedefaultseppunct}\relax
\EndOfBibitem
\bibitem[Szamel(2014)]{szamel2014self}
G.~Szamel, \emph{Physical Review E}, 2014, \textbf{90}, 012111\relax
\mciteBstWouldAddEndPuncttrue
\mciteSetBstMidEndSepPunct{\mcitedefaultmidpunct}
{\mcitedefaultendpunct}{\mcitedefaultseppunct}\relax
\EndOfBibitem
\bibitem[Mijalkov and Volpe(2013)]{mijalkov2013sorting}
M.~Mijalkov and G.~Volpe, \emph{Soft Matter}, 2013, \textbf{9},
  6376--6381\relax
\mciteBstWouldAddEndPuncttrue
\mciteSetBstMidEndSepPunct{\mcitedefaultmidpunct}
{\mcitedefaultendpunct}{\mcitedefaultseppunct}\relax
\EndOfBibitem
\bibitem[Volpe \emph{et~al.}(2014)Volpe, Gigan, and Volpe]{volpe2014simulation}
G.~Volpe, S.~Gigan and G.~Volpe, \emph{American Journal of Physics}, 2014,
  \textbf{82}, 659--664\relax
\mciteBstWouldAddEndPuncttrue
\mciteSetBstMidEndSepPunct{\mcitedefaultmidpunct}
{\mcitedefaultendpunct}{\mcitedefaultseppunct}\relax
\EndOfBibitem
\bibitem[Li \emph{et~al.}(2014)Li, Ghosh, Marchesoni, and
  Li]{li2014manipulating}
Y.~Li, P.~K. Ghosh, F.~Marchesoni and B.~Li, \emph{Physical Review E}, 2014,
  \textbf{90}, 062301\relax
\mciteBstWouldAddEndPuncttrue
\mciteSetBstMidEndSepPunct{\mcitedefaultmidpunct}
{\mcitedefaultendpunct}{\mcitedefaultseppunct}\relax
\EndOfBibitem
\bibitem[Ai \emph{et~al.}(2015)Ai, He, and Zhong]{ai2015chirality}
B.-q. Ai, Y.-f. He and W.-r. Zhong, \emph{Soft Matter}, 2015, \textbf{11},
  3852--3859\relax
\mciteBstWouldAddEndPuncttrue
\mciteSetBstMidEndSepPunct{\mcitedefaultmidpunct}
{\mcitedefaultendpunct}{\mcitedefaultseppunct}\relax
\EndOfBibitem
\bibitem[Ao \emph{et~al.}(2015)Ao, Ghosh, Li, Schmid, H{\"a}nggi, and
  Marchesoni]{ao2015diffusion}
X.~Ao, P.~K. Ghosh, Y.~Li, G.~Schmid, P.~H{\"a}nggi and F.~Marchesoni,
  \emph{EPL (Europhysics Letters)}, 2015, \textbf{109}, 10003\relax
\mciteBstWouldAddEndPuncttrue
\mciteSetBstMidEndSepPunct{\mcitedefaultmidpunct}
{\mcitedefaultendpunct}{\mcitedefaultseppunct}\relax
\EndOfBibitem
\bibitem[Marconi and Maggi(2015)]{marconi2015towards}
U.~M.~B. Marconi and C.~Maggi, \emph{Soft matter}, 2015, \textbf{11},
  8768--8781\relax
\mciteBstWouldAddEndPuncttrue
\mciteSetBstMidEndSepPunct{\mcitedefaultmidpunct}
{\mcitedefaultendpunct}{\mcitedefaultseppunct}\relax
\EndOfBibitem
\bibitem[Das \emph{et~al.}(2018)Das, Gompper, and Winkler]{das2018confined}
S.~Das, G.~Gompper and R.~G. Winkler, \emph{New Journal of Physics}, 2018,
  \textbf{20}, 015001\relax
\mciteBstWouldAddEndPuncttrue
\mciteSetBstMidEndSepPunct{\mcitedefaultmidpunct}
{\mcitedefaultendpunct}{\mcitedefaultseppunct}\relax
\EndOfBibitem
\bibitem[Caprini \emph{et~al.}(2018)Caprini, Marconi, and
  Puglisi]{caprini2018activity}
L.~Caprini, U.~M.~B. Marconi and A.~Puglisi, \emph{arXiv preprint
  arXiv:1810.12652}, 2018\relax
\mciteBstWouldAddEndPuncttrue
\mciteSetBstMidEndSepPunct{\mcitedefaultmidpunct}
{\mcitedefaultendpunct}{\mcitedefaultseppunct}\relax
\EndOfBibitem
\bibitem[Fily and Marchetti(2012)]{fily2012athermal}
Y.~Fily and M.~C. Marchetti, \emph{Physical Review Letters}, 2012,
  \textbf{108}, 235702\relax
\mciteBstWouldAddEndPuncttrue
\mciteSetBstMidEndSepPunct{\mcitedefaultmidpunct}
{\mcitedefaultendpunct}{\mcitedefaultseppunct}\relax
\EndOfBibitem
\bibitem[Farage \emph{et~al.}(2015)Farage, Krinninger, and
  Brader]{farage2015effective}
T.~F. Farage, P.~Krinninger and J.~M. Brader, \emph{Physical Review E}, 2015,
  \textbf{91}, 042310\relax
\mciteBstWouldAddEndPuncttrue
\mciteSetBstMidEndSepPunct{\mcitedefaultmidpunct}
{\mcitedefaultendpunct}{\mcitedefaultseppunct}\relax
\EndOfBibitem
\bibitem[Marconi \emph{et~al.}(2016)Marconi, Gnan, Paoluzzi, Maggi, and
  Di~Leonardo]{marconi2016velocity}
U.~M.~B. Marconi, N.~Gnan, M.~Paoluzzi, C.~Maggi and R.~Di~Leonardo,
  \emph{Scientific reports}, 2016, \textbf{6}, 23297\relax
\mciteBstWouldAddEndPuncttrue
\mciteSetBstMidEndSepPunct{\mcitedefaultmidpunct}
{\mcitedefaultendpunct}{\mcitedefaultseppunct}\relax
\EndOfBibitem
\bibitem[Caprini and Marconi(2018)]{caprini2018active}
L.~Caprini and U.~M.~B. Marconi, \emph{Soft matter}, 2018\relax
\mciteBstWouldAddEndPuncttrue
\mciteSetBstMidEndSepPunct{\mcitedefaultmidpunct}
{\mcitedefaultendpunct}{\mcitedefaultseppunct}\relax
\EndOfBibitem
\bibitem[Howse \emph{et~al.}(2007)Howse, Jones, Ryan, Gough, Vafabakhsh, and
  Golestanian]{howse2007self}
J.~R. Howse, R.~A. Jones, A.~J. Ryan, T.~Gough, R.~Vafabakhsh and
  R.~Golestanian, \emph{Physical review letters}, 2007, \textbf{99},
  048102\relax
\mciteBstWouldAddEndPuncttrue
\mciteSetBstMidEndSepPunct{\mcitedefaultmidpunct}
{\mcitedefaultendpunct}{\mcitedefaultseppunct}\relax
\EndOfBibitem
\bibitem[Risken(1984)]{risken}
H.~Risken, \emph{Fokker-Planck Equation}, Springer, 1984\relax
\mciteBstWouldAddEndPuncttrue
\mciteSetBstMidEndSepPunct{\mcitedefaultmidpunct}
{\mcitedefaultendpunct}{\mcitedefaultseppunct}\relax
\EndOfBibitem
\bibitem[Puglisi and Marini Bettolo~Marconi(2017)]{puglisi2017clausius}
A.~Puglisi and U.~Marini Bettolo~Marconi, \emph{Entropy}, 2017, \textbf{19},
  356\relax
\mciteBstWouldAddEndPuncttrue
\mciteSetBstMidEndSepPunct{\mcitedefaultmidpunct}
{\mcitedefaultendpunct}{\mcitedefaultseppunct}\relax
\EndOfBibitem
\bibitem[Cates(2012)]{cates2012diffusive}
M.~Cates, \emph{Reports on Progress in Physics}, 2012, \textbf{75},
  042601\relax
\mciteBstWouldAddEndPuncttrue
\mciteSetBstMidEndSepPunct{\mcitedefaultmidpunct}
{\mcitedefaultendpunct}{\mcitedefaultseppunct}\relax
\EndOfBibitem
\bibitem[Marconi \emph{et~al.}(2017)Marconi, Puglisi, and
  Maggi]{marconi2017heat}
U.~M.~B. Marconi, A.~Puglisi and C.~Maggi, \emph{Scientific reports}, 2017,
  \textbf{7}, 46496\relax
\mciteBstWouldAddEndPuncttrue
\mciteSetBstMidEndSepPunct{\mcitedefaultmidpunct}
{\mcitedefaultendpunct}{\mcitedefaultseppunct}\relax
\EndOfBibitem
\bibitem[ten Hagen \emph{et~al.}(2011)ten Hagen, van Teeffelen, and
  L{\"o}wen]{ten2011brownian}
B.~ten Hagen, S.~van Teeffelen and H.~L{\"o}wen, \emph{Journal of Physics:
  Condensed Matter}, 2011, \textbf{23}, 194119\relax
\mciteBstWouldAddEndPuncttrue
\mciteSetBstMidEndSepPunct{\mcitedefaultmidpunct}
{\mcitedefaultendpunct}{\mcitedefaultseppunct}\relax
\EndOfBibitem
\bibitem[Dauchot and D{\'e}mery(2018)]{dauchot2018dynamics}
O.~Dauchot and V.~D{\'e}mery, \emph{arXiv preprint arXiv:1810.13303},
  2018\relax
\mciteBstWouldAddEndPuncttrue
\mciteSetBstMidEndSepPunct{\mcitedefaultmidpunct}
{\mcitedefaultendpunct}{\mcitedefaultseppunct}\relax
\EndOfBibitem
\bibitem[Hanggi and Jung(1995)]{hanggi1995colored}
P.~Hanggi and P.~Jung, \emph{Advances in Chemical Physics}, 1995, \textbf{89},
  239--326\relax
\mciteBstWouldAddEndPuncttrue
\mciteSetBstMidEndSepPunct{\mcitedefaultmidpunct}
{\mcitedefaultendpunct}{\mcitedefaultseppunct}\relax
\EndOfBibitem
\bibitem[Marconi \emph{et~al.}(2016)Marconi, Maggi, and
  Melchionna]{marconi2016pressure}
U.~M.~B. Marconi, C.~Maggi and S.~Melchionna, \emph{Soft matter}, 2016,
  \textbf{12}, 5727--5738\relax
\mciteBstWouldAddEndPuncttrue
\mciteSetBstMidEndSepPunct{\mcitedefaultmidpunct}
{\mcitedefaultendpunct}{\mcitedefaultseppunct}\relax
\EndOfBibitem
\bibitem[Marini Bettolo~Marconi \emph{et~al.}(2017)Marini Bettolo~Marconi,
  Maggi, and Paoluzzi]{marini2017pressure}
U.~Marini Bettolo~Marconi, C.~Maggi and M.~Paoluzzi, \emph{The Journal of
  chemical physics}, 2017, \textbf{147}, 024903\relax
\mciteBstWouldAddEndPuncttrue
\mciteSetBstMidEndSepPunct{\mcitedefaultmidpunct}
{\mcitedefaultendpunct}{\mcitedefaultseppunct}\relax
\EndOfBibitem
\bibitem[Wensink and L{\"o}wen(2008)]{wensink2008aggregation}
H.~Wensink and H.~L{\"o}wen, \emph{Physical Review E}, 2008, \textbf{78},
  031409\relax
\mciteBstWouldAddEndPuncttrue
\mciteSetBstMidEndSepPunct{\mcitedefaultmidpunct}
{\mcitedefaultendpunct}{\mcitedefaultseppunct}\relax
\EndOfBibitem
\bibitem[Elgeti and Gompper(2013)]{elgeti2013wall}
J.~Elgeti and G.~Gompper, \emph{EPL (Europhysics Letters)}, 2013, \textbf{101},
  48003\relax
\mciteBstWouldAddEndPuncttrue
\mciteSetBstMidEndSepPunct{\mcitedefaultmidpunct}
{\mcitedefaultendpunct}{\mcitedefaultseppunct}\relax
\EndOfBibitem
\bibitem[Lee(2013)]{lee2013active}
C.~F. Lee, \emph{New Journal of Physics}, 2013, \textbf{15}, 055007\relax
\mciteBstWouldAddEndPuncttrue
\mciteSetBstMidEndSepPunct{\mcitedefaultmidpunct}
{\mcitedefaultendpunct}{\mcitedefaultseppunct}\relax
\EndOfBibitem
\bibitem[Liao and Klapp(2018)]{liao2018clustering}
G.-J. Liao and S.~H. Klapp, \emph{Soft matter}, 2018, \textbf{14},
  7873--7882\relax
\mciteBstWouldAddEndPuncttrue
\mciteSetBstMidEndSepPunct{\mcitedefaultmidpunct}
{\mcitedefaultendpunct}{\mcitedefaultseppunct}\relax
\EndOfBibitem
\bibitem[Meinhardt \emph{et~al.}(2012)Meinhardt, Smiatek, Eichhorn, and
  Schmid]{meinhardt2012separation}
S.~Meinhardt, J.~Smiatek, R.~Eichhorn and F.~Schmid, \emph{Physical review
  letters}, 2012, \textbf{108}, 214504\relax
\mciteBstWouldAddEndPuncttrue
\mciteSetBstMidEndSepPunct{\mcitedefaultmidpunct}
{\mcitedefaultendpunct}{\mcitedefaultseppunct}\relax
\EndOfBibitem
\end{mcitethebibliography}

\end{document}